\documentclass[prb,aps,amsmath,amssymb,twocolumn,superscriptaddress,10pt]{revtex4-2}
\usepackage{hyperref}
\hypersetup{
breaklinks=true,
colorlinks=true,
citecolor=blue,
linkcolor=magenta,
filecolor=blue,
urlcolor=blue
}
\usepackage{graphicx}

\begin{document}


\title{Acoustic Full Waveform Inversion with Hamiltonian Monte Carlo Method}
\author{Paulo Douglas Santos de Lima}
\affiliation{Departamento de Física Teórica e Experimental, Universidade Federal do Rio Grande do Norte, 59078-970 Natal-RN, Brazil}%

\author{Gilberto Corso}
\affiliation{Departamento de Biofísica e Farmacologia, \\ Universidade Federal do Rio Grande do Norte, 59078-970 Natal-RN, Brazil}%

\author{M. S. Ferreira}
\affiliation{School of Physics, Trinity College Dublin, Dublin 2, Ireland}
\affiliation{Centre for Research on Adaptive Nanostructures and Nanodevices (CRANN) \& Advanced Materials and Bioengineering Research (AMBER) Centre, Trinity College Dublin, Dublin 2, Ireland}

\author{João Medeiros de Araújo}%
\affiliation{Departamento de Física Teórica e Experimental, Universidade Federal do Rio Grande do Norte, 59078-970 Natal-RN, Brazil}%

\date{\today}

\begin{abstract}
Full-Waveform Inversion (FWI) is a high-resolution technique used in geophysics to evaluate the physical parameters and construct subsurface models in a noisy and limited data scenario. The ill-posed nature of the FWI turns this a challenging problem since more than one model can match the observations. In a probabilistic way, solving the FWI problem demands efficient sampling techniques to infer information on parameters and to estimate the uncertainties in high-dimensional model spaces. We investigate the feasibility of applying the Hamiltonian Monte Carlo (HMC) method in the acoustic FWI by a reflection setup containing different noise level data. We propose a new strategy for tuning the mass matrix based on the acquisition geometry of the seismic survey. Our methodology significantly improves the ability of the HMC method in reconstructing reasonable seismic models with affordable computational efforts.
\end{abstract}

\maketitle


\section{\label{sec:level1}Introduction}

The problem of constructing consistent physical models of the Earth’s subsurface based on observations of the complete seismic-wave propagation is named Full Waveform Inversion (FWI). Despite being developed in the late 70's~\cite{tarantola1984}, this method has become particularly useful in the past decade due to the impressive advances in the computational power of modern devices in tandem with some ingenious numerical modelling techniques that are now available \cite{pratt1990}.

FWI is a natural extension of travel-time tomography~\cite{aki77} in which  not only the phase information is recorded but also the amplitude, providing better resolution of the subsurface when compared to standard methods. 
FWI is a nonlinear and ill-posed problem in which the physical parameters (e.g density, velocity) are estimated from an information source that is limited in space and frequency and is more often than not in the presence of heavy noise~\cite{gras2019, fruehn2019}. Moreover, inaccurate modeling and parametrization methods combined with insufficient prior knowledge of the system are also factors that introduce uncertainties into the inversion results~\cite{sen_stoffa_2013}. All these limitations make FWI a particularly challenging problem since in practice solutions are not necessarily unique. Therefore, quantifying the uncertainty of results, {\it i.e.} how believable they are, is a fundamental task in FWI, mainly in oil and gas exploration. The reliability of results can be used to assert features of the subsurface that are well resolved or even if more field data needs to be collected~\cite{osypov2013, raw2014}.



The basis of this inversion problem consists in minimising the difference between the observed and the modelled data, which is called the residuals.  Two main strategies are used in the FWI optimization process: the deterministic and the probabilistic.  In the deterministic approach an initial condition is evolved according to a dynamical rule in order to find a minimum of the residuals.  The deterministic method relies on algorithms based on gradient of the error function with respect to the model parameters~\cite{brossier2009,metivier2014}. However, while these methods provide a single inverted solution that is minimally deviated from the observed data, they offer no information about the uncertainty of the physical parameters~\cite{Tromp2020}.




In contrast to the deterministic strategy, in the stochastic version of the FWI problem, the solution is treated as a probability distribution and requires the use of efficient sampling techniques~\cite{hoang2013complexity}. Markov Chain Monte Carlo (MCMC) is the most commonly used tool for this task, where the inversion result is expressed in terms of the mean, variance and/or other statistically relevant quantities~\cite{neal2011}. Nevertheless, the MCMC is inefficient to estimate probability distributions in high dimensional model spaces, which is the typical scenario in seismic inversion. 
This inefficiency occurs due to the so-called curse of dimensionality, which asserts that the number of relevant models decreases rapidly with increasing  model space dimension. The Hamiltonian Monte Carlo~\cite{DUANE1987216, betancourt2018conceptual} (HMC) method is a potentially good candidate to overcome this drawback. It contains the gradient information present in local optimization methods together with the flexibility of the derivative-free MCMC methods. 
In this way, the HMC is a hybrid method that attempts to combine the best of the deterministic and probabilistic approaches using a deterministic exploration of particular level sets of energy but with stochastic exploration among them.

Originally, the HMC was formulated to be applied in quantum chromodynamics, but it has now been implemented in neural networks, machine learning~\cite{neal1996}, molecular simulations~\cite{dubbeldam2016} and quantum mechanics~\cite{shang2015}, to name but a few. Recently, this approach has been popularized in geophysical applications, as for example in amplitude versus angle inversion~\cite{Aleardi2020}, seismic point source inversion~\cite{Fichtner2018}, elastic FWI~\cite{Gebraad2020} and extensions of original HMC method using reversible jumps~\cite{biswas2017, Aleardi2020b}. However, applications of HMC in complex seismic models in reflection setup have been underdeveloped to date, most likely due to the difficulty of adequately choosing and tuning the particle masses, which has a crucial role in sampling the canonical distribution of Hamiltonian systems. In fact, the success of HMC in nonlinear inverse problems is strongly dependent on the existence of a suitable mass matrix that allows efficient exploration of the phase space~\cite{fichtner2021}.


The implementation of the HMC method in acoustic FWI requires great numerical care and while there is plenty of room for improvements in that area, this is not our primary goal. At this stage, our focus is on the feasibility of a HMC-based approach to the FWI problem within the acoustic wave approximation. At the heart of the HMC method is the tuning of a few free parameters that can speed up the finding of global minima in the FWI problem and in turn alleviate some of the effects caused by the curse of dimensionality. With that in mind, we aim to identify an appropriate methodology to select and tune the effective mass of the Hamiltonian dynamics which, as we shall see, have a mathematical interpretation in the search algorithm. The remainder of the paper is organized as follows: in section II we sketch the FWI method, in section III we show in some detail the HMC strategy, in section IV the numerical experiment is exposed, in section V the results are outlined, and finally in section VI we present the main conclusion of the work.

\section{Theory}
\subsection{Full Waveform Inversion}

FWI is specified by three main ingredients: (1) the seismic wavefield observations, (2) the physical properties of the subsurface that we wish to describe and (3) the (nonlinear) theory that relates the observations with the physical properties. The first two ingredients are encapsulated in the observed data, hereafter represented by the quantity $\mathbf{d}^{obs}$, and the model vectors $\mathbf{m}$. It is worth highlighting that the modelled data vector $\mathbf{d}^{mod}$ is constructed from the model in order to compare the predictions with observations through the residuals $\Delta\mathbf{d} = \mathbf{d}^{mod} - \mathbf{d}^{obs}$. The set of plausible models and the data obtained from it span the model $\mathbb{M}$ and the data $\mathbb{D}$ spaces, respectively.

We consider that the subsurface is approximated by a two-dimensional acoustic medium~\cite{tarantola84} with spatial coordinates $\mathbf{x} = (x, z)$, where $x$ and $z$ are the horizontal distance and the depth of the model. Following this assumption, to compute the modelled data we first define the acquisition geometry, that is, the number of sources $N_s$ and receivers $N_r$ and their respective positions $\{\mathbf{x}_s\}_{s= 1,\ldots,N_s}$ and $\{\mathbf{x}_r\}_{r= 1,\ldots,N_r}$. We denote the coordinates of residuals by $\Delta d_{r,s}(\mathbf{m}, t)$ to emphasize the source/receiver dependence.

The connection between the observations and the physical properties of the subsurface is obviously captured by the acoustic wave equation:
\begin{equation}
    \nabla^2u_s(\mathbf{x}, t) -m(\mathbf{x})\frac{\partial^2 u_s(\mathbf{x}, t)}{\partial t^2} = s(t)\delta(\mathbf{x}-\mathbf{x}_s)\,, 
\label{wave_eq}
\end{equation}
where $u_s(\mathbf{x}, t)$ is the time-dependent seismic wavefield probed at the receiver position $\mathbf{x}$ as a response to the acoustic excitation $s(t)$ generated by a given source $s$. In this approach, the coefficients of square slowness $m(\mathbf{x}) = \nu(\mathbf{x})^{-2}$ (where $\nu$ is the acoustic velocity) expanded in a regular basis of spatial domain are called model parameters $m_i$ and constitutes the model $\mathbf{m}$. 

For simplicity we use the Ricker wavelet~\cite{wang2014, wang2015} as seismic source: 
\begin{equation}
    s(t) = (1 - 2\pi^2f_0^2t^2)\exp{(-\pi^2 f_0^2t^2)}\,,
    \label{ricker}
\end{equation}
where $f_0$ is the central frequency.

In the probabilistic point of view, the modelled data is interpreted as a random vector and the probability that a proposed model $\mathbf{m}$ explains the observed data $\mathbf{d}^{obs}$ is given by the likelihood function $\mathcal{L}(\mathbf{m}) \propto \exp{(-E(\mathbf{m}))}$, which compares the modelled and observed data through some misfit function~\cite{tarantola2005}. We assume an uncorrelated Gaussian-distributed data, such that the misfit is written as:
\begin{equation}
    E(\mathbf{m}) = \frac{1}{2}\Delta\mathbf{d}^\text{T}(\mathbf{m})\Sigma^{-1}\Delta\mathbf{d}(\mathbf{m})\,, \label{misfit_function}
\end{equation}
where $\Sigma$ is the noise covariance matrix, which we choose to be $\Sigma = \sigma^2\mathbf{I}$. For our synthetic study, the variance of residuals $\sigma^2$ is assumed known and can be considered as a fixed parameter during the inversion~\cite{sambridge2013}. Despite the normality about the residuals distribution, we make no assumption about the model distribution. It is worth mentioning that the level of imprecision contained in real observation data depends crucially on the seismic surveys and therefore it is paramount to be able to estimate the level of uncertainty contained in the data~\cite{bodin2012} combined with other suitable misfit functions~\cite{metivier2016, liu2016, carvalho2021}.

\subsection{Hamiltonian Monte Carlo}

In the HMC method the model parameters $\mathbf{m}$ are interpreted as a set of particles moving along trajectories of a classical mechanical system. The particles have effective mass $\mu$ and are subjected to an artificial potential energy that mimics the misfit function defined in Eq. \eqref{misfit_function}. Bearing in mind that the FWI consists in searching the model parameters that minimise the misfit function, it is understandable why we establish a parallel with a mechanical system whose dynamics naturally evolve to minimise its total energy.  Accordingly, the model space $\mathbb{M}$ is extended to a (fake) phase space $\mathbb{Z} = \mathbb{M}\times\mathbb{P}$, such that the  likelihood function is obtained through sampling over the canonical distribution:  
\begin{equation}
    \rho(\mathbf{m}, \mathbf{p}) \propto \exp{(-H(\mathbf{m}, \mathbf{p}))}\,,
\label{can_dist}
\end{equation}
with a Hamiltonian $H(\mathbf{m}, \mathbf{p})$ given by:
\begin{equation}
    H(\mathbf{m}, \mathbf{p}) = \frac{1}{2}\mathbf{p}^\text{T}\mathbf{M}^{-1}\mathbf{p} + E(\mathbf{m})\,.
\label{hamiltonian}
\end{equation}

In the equation above we chose a simple form for the kinetic term where the momenta $\mathbf{p} \in \mathbb{P}$ is randomly sampled according to a Gaussian distribution with zero mean and a covariance given by the (diagonal) matrix mass $\mathbf{M} = \mu\mathbf{I}$, which is an important parameter of the HMC numerical simulations.

To sample the distribution \eqref{can_dist}, we first evolve over (artificial) time $\tau$ an initial state $(\mathbf{m}_0, \mathbf{p}_0)$ using  the dynamics of  Hamilton equations~\cite{lemos_2018}: 
\begin{equation}
    \frac{d\mathbf{m}}{d\tau} = \frac{\mathbf{p}}{\mu}\,, \quad \frac{d\mathbf{p}}{d\tau} = -\nabla E(\mathbf{m})\,.
    \label{hamiltons_eq2}
\end{equation}

In the sequence, the final state $(\mathbf{m}_\tau, \mathbf{p}_\tau)$  is accepted with a probability given by the Metropolis-Hasting criteria~\cite{metropolis1953}: 
\begin{equation}
    \mathrm{min}\left[1, \exp{(H(\mathbf{m}_0, \mathbf{p}_0)-H \left(\mathbf{m}_\tau, \mathbf{p}_\tau)\right)}\right]\,.
    \label{MH_criteria}
\end{equation}

When Eq. \eqref{MH_criteria} is satisfied $\mathbf{m}_\tau$ is stored as a sample model and $\mathbf{p}_\tau$ is discarded. Subsequently, $\mathbf{m}_\tau$ is employed as the new initial position that is again evolved by the dynamical Eqs. (\ref{hamiltons_eq2}) with a brand new set of random momenta. This procedure is referred to as a single HMC step, which is then repeated $N_{\text{HMC}}$ times that produce $N$ samples of the canonical distribution \eqref{can_dist}. In fact, only a fraction of samples are accepted and thus we use the acceptance rate $r = N/N_{\text{HMC}}$ of the samples as a control parameter to tune the HMC method~\cite{leimkuhler94}. In this spirit, we adjust the HMC parameters (Sec. \ref{hmc_parameters}) trying to maintain high values ($r > 0.6$) of acceptance rate. At the end of a HMC simulation, the set of $N$ samples are used to quantify the uncertainty in the result, by looking at the sample statistical moments for the acoustic velocity.

\section{Numerical Experiments}

\begin{figure*}[htbp]
    \centering
    \includegraphics[width=6.8in]{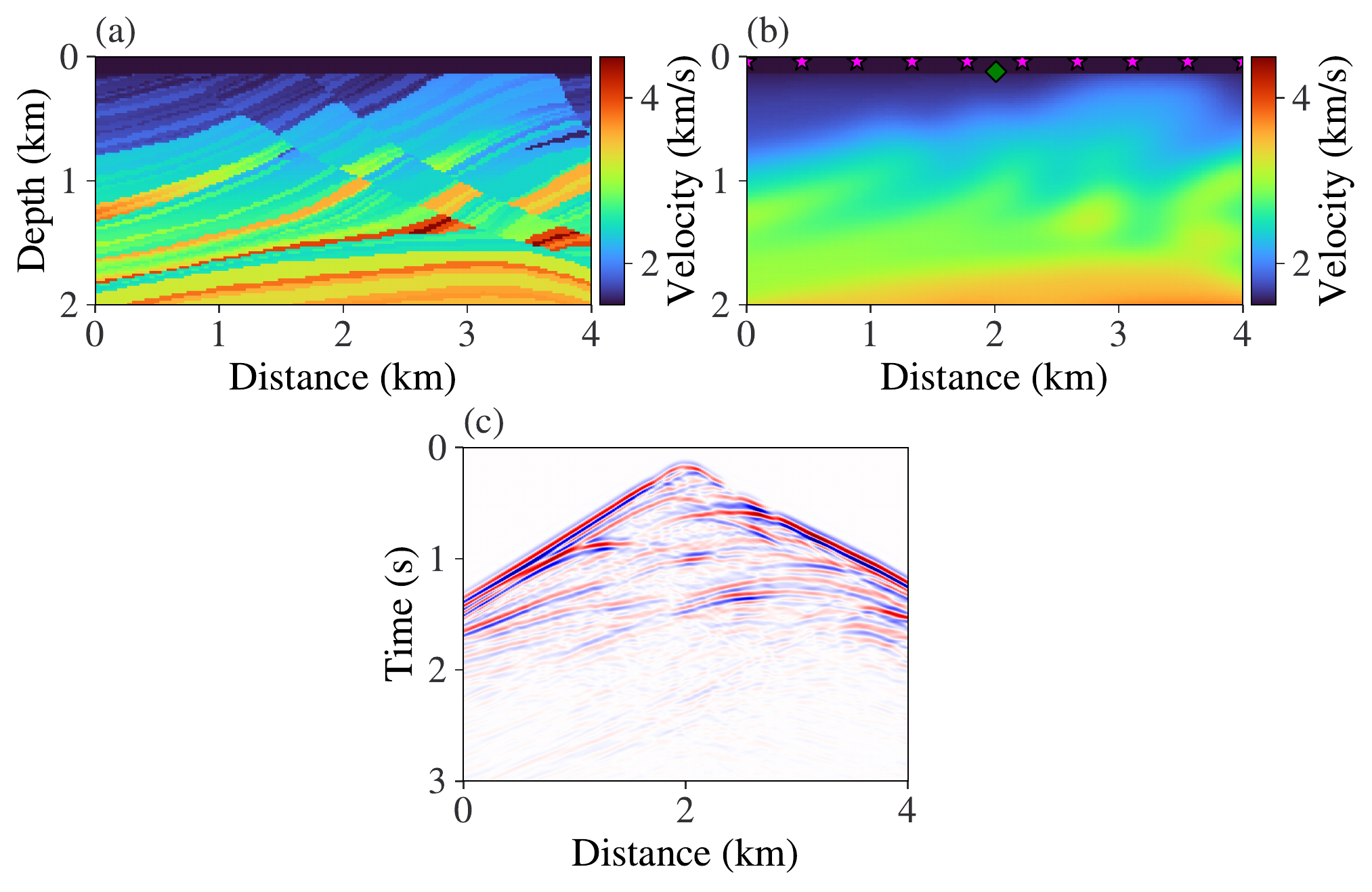}
    \caption{The Marmousi velocity model (a) possess a geometry with abrupt velocity variations from 1.5 km/s to 4.5 km/s. The initial model (b) is a smoother version of target model (a). The purple stars denote the locations of 10 sources and the green square indicates the position of one of 200 receivers in (b). An example of seismogram (for a single shot located at $x = 2.0$ km and $z = 40$ m) (c) which shows the residuals computed from all receivers during the time recording of the initial model (b).}
    \label{fig:marm_target_init}
\end{figure*}

We test the feasibility of combining HMC and FWI with a cropped version of the Marmousi model (Fig. \ref{fig:marm_target_init}a) which is based on the geology of the Kwanza basin region in Angola~\cite{versteeg1994} and is widely used as a benchmark model in seismic inversion~\cite{martin2006}. Our simulations represent a maritime reflection seismic experiment~\cite{wencai2013reflection}, meaning that the sources and receivers are placed in the water layer (top of Marmousi model), which is assumed to have a constant velocity of 1.5 km/s throughout the inversion procedure. We use the model presented in Fig. \ref{fig:marm_target_init}b as the initial position $\mathbf{m}_0$ for our HMC simulations. At this point, it is important to emphasize that all information used during the inversion is contained in data residuals, which can be visualized in the seismograms as illustrated in Fig. \ref{fig:marm_target_init}c.

The domain-specific language DEVITO~\cite{devito-compiler, devito-api} was used for simulating the acoustic wave propagation using a finite difference approximation scheme with eight-order spatial derivatives and second-order time derivatives. Further details concerning applications in seismic modelling with DEVITO can be founded in \cite{devito_tuto} (and references therein). The velocity model was discretized in a $281\times 156$ regular grid, yielding a 43836-dimensional model space. In addition, an infinite domain was mimicked with a damping term in \eqref{wave_eq} to attenuate the wavefield outside the simulation boundaries and avoid unphysical reflection during the simulations~\cite{clayton77}. 

The data set was generated using $10$ sources \eqref{ricker} with a mean frequency equal to $10$ Hz, which are located at every 400 m and at 40 m depth. The data acquisition was realized during $5$ s by $200$ receivers located every 20 m, deployed at 120 m depth. Bearing in mind that the noise in the data affects the resultant seismic models, we investigate the robustness of the HMC method by simulating a high ($\sigma^2 = 10$), medium ($\sigma^2 = 1.0$) and low ($\sigma^2 = 0.1$) noise scenarios.

\subsection{Leapfrog Integration and Gradient Calculation}\label{hmc_parameters}

Numerical errors associated with the Hamiltonian dynamics \eqref{hamiltons_eq2} simulation impair the energy conservation which diminishes the model acceptance in \eqref{MH_criteria}. Fortunately, other properties of Hamiltonian systems such as time reversibility and volume preservation are protected when a symplectic integrator is employed. For this reason, we opt for the leapfrog method, which has a symplectic nature and discretizes Hamilton equations in $L$ leapfrog steps of size $\epsilon$ with global error $\mathcal{O}(\epsilon^2)$. We implement a modified version of this method that considers prior knowledge on the acceptable minimum and maximum seismic velocities for the proposed models~\cite{neal2011}: 1.5 km/s and 4.5 km/s. Although we are not following a Bayesian approach, we stress that the samples produced from the initial model $\mathbf{m}_0$ combined with this velocity bounds can be seen as a uniform prior distribution used in the generation of the samples.


The computation of  the gradient in \eqref{hamiltons_eq2} is the most demanding task of the FWI workflow. This cost is mitigated by using the adjoint state method~\cite{plessix2006, virieux2009}, which replaces the Jacobian calculation by an additional wave propagation. This method constructs the gradient (subsurface imaging) by crosscorrelating the second time derivative of the seismic wavefield $u_s(\mathbf{x}, t)$ with the adjoint wavefield $v_s(\mathbf{x}, t)$, the latter being achieved by backpropagating (in time) the seismic wavefield using $f_s(t) = -\frac{1}{\sigma}\sum_{r=1}^{N_r}\Delta d_{s, r}(\mathbf{m}, T-t)\delta(\mathbf{x}-\mathbf{x}_r)$ as the (adjoint) source term~\cite{tromp2005}. Therefore, the gradient can be written as
\begin{equation}
    \nabla E(\mathbf{m}) = -\sum_{s=1}^{N_s}\int_0^{T} v_s(\mathbf{x}, T - t)\frac{\partial^2u_s(\mathbf{x}, t)}{\partial t^2}\,dt\,, \label{grad_adj}
\end{equation}
being discretized during the simulations following the aforementioned finite difference scheme.

\subsection{Tuning HMC Parameters}

Sampling the canonical distribution \eqref{can_dist} using HMC involves a careful tuning of $L$, $\epsilon$ and $\mathbf{M}$ to effectively explore the phase space and, in turn, bring computational gains. The computational cost is mainly due to the gradient \eqref{grad_adj} which must be calculated $2L$ times for each Hamiltonian trajectory of length $L\epsilon$. In this way, we first tune $L$ and $\epsilon$ considering that long trajectories can be associated to particles that visit the same region of phase space several times while short trajectories may be associated to particles that remain near the initial position. After some preliminary tests, we fixed these values as $L = 5$ and $\epsilon = 10^{-3}$.

\begin{figure*}[htbp]
    \centering
    \includegraphics[width=6.8in]{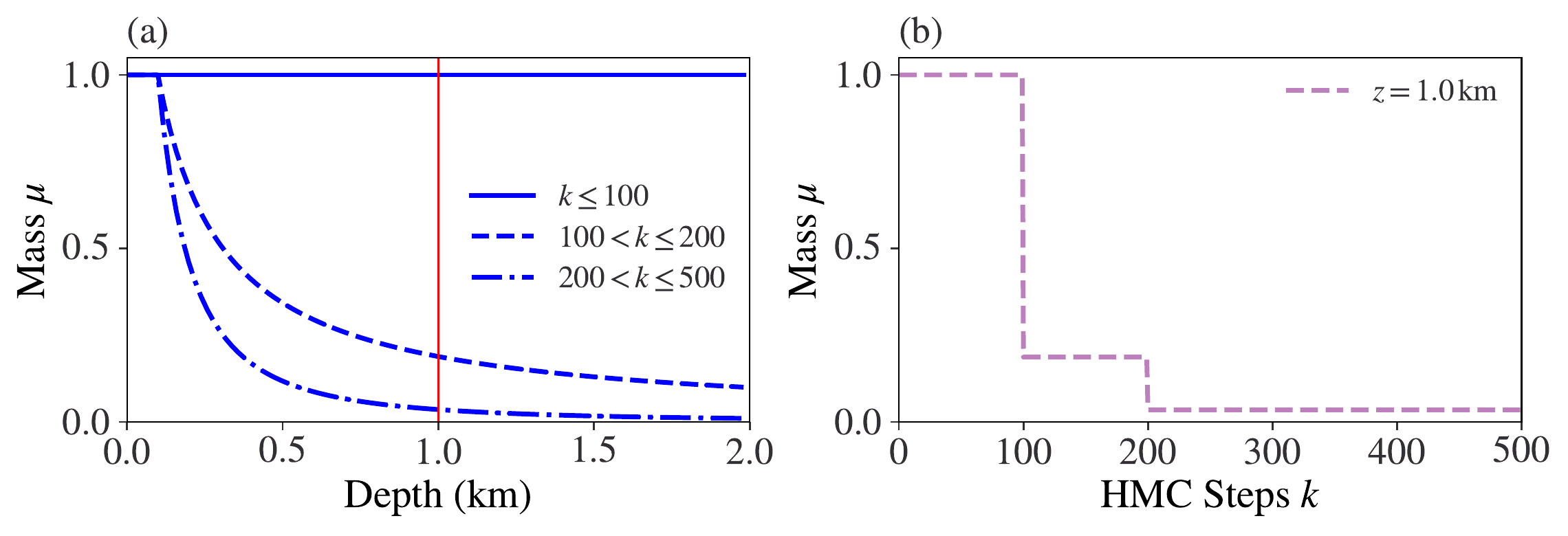}
    \caption{Tuning strategy of the matrix mass for the high noise case. (a) The particle mass varies according to depth and the number $k$ of HMC steps. (b) The mass values to particles located at depth $z = 1.0$ km (red line in Fig. \ref{fig:tuning_mass}a) during the HMC exploration.}
    \label{fig:tuning_mass}
\end{figure*}

In contrast with $L$ and $\epsilon$, the mass matrix $\mathbf{M}$ can be tuned according to the seismic velocities in the subsurface.  
We propose a new strategy based on the lack of information with depth in reflection seismic experiments. Firstly, we attribute the same mass $\mu$ to each model parameter $m_i$ and, after a certain number of HMC steps (i.e some phase space exploration), each particle mass is divided by a monotonically increasing function $\gamma_i(z)$ that depends on the depth $z$ in the seismic model. Physically, this corresponds to making the particles lighter as the system gets close to a minimum of potential energy, which is sensitive to model depth because of the acquisition geometry. Although this procedure can in principle be executed repeatedly, in our case there is no need to do it more than twice. We emphasize that model parameters located at the same depth but with different horizontal positions will always possess the same mass. 

Our tuning strategy is illustrated in Figure \ref{fig:tuning_mass} to the Marmousi inversion. We use $N_{\text{HMC}} = 500$ and initial masses of $\mu = 1.0$ which are diminished every 100 HMC steps using
\begin{equation}
    \gamma_i(z) = \gamma_{\text{min}} + (\gamma_{\text{max}} - \gamma_{\text{min}})\left(\frac{z - z_{\text{water}}}{z_{\text{max}} - z_{\text{water}}}\right)\,,
\end{equation}
where $z_{\text{water}} \leq z \leq z_{\text{max}}$ and $z_{\text{max}} = 2$ km is the model depth, $z_{\text{water}} = 0.12$ km is the water layer depth. We have tested several values for $\gamma_{\text{min}}$ and $\gamma_{\text{max}}$, but the better results were obtained when we set $\gamma_{\text{min}} = 1.0$ and $\gamma_{\text{max}} = 1.5, 7.5$ and $10$ for $\sigma^2 = 0.1, 1.0$ and $10$, respectively. The conventional choice for the matrix mass is recovered by setting $\gamma_{\text{max}} = 1.0$.

\section{Results}

\begin{figure*}[htbp]
    \centering
    \includegraphics[width=6.8in]{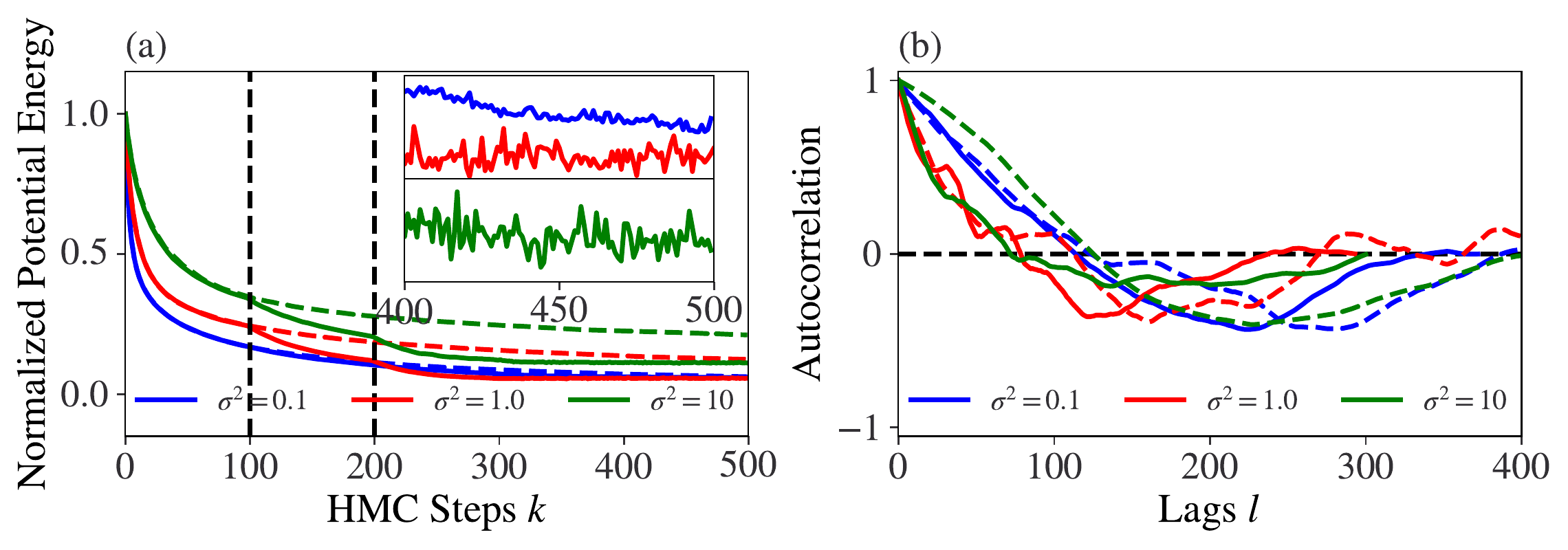}
    \caption{Convergence analysis for noise scenarios: (a) normalized potential energy (misfit function) as a function of the HMC steps and (b) an example of the autocorrelation for the velocity parameter no. 16356 (located at $x = 2$ km and $z = 1.5$ km). The colored solid lines point out our tuning strategy for the mass matrix while the dashed one represents the standard choice $\mathbf{M} = \mathbf{I}$ and the vertical dashed lines in (a) indicate the moments when the masses are reduced. The HMC convergence is directly affected by the data noise and it is speeds up by our strategy.}
    \label{fig:convergence}
\end{figure*}

Figure \ref{fig:convergence} shows the effect of tuning the matrix mass in the HMC convergence in comparison with the standard choice for the mass matrix (fixed mass matrix $\mathbf{M} = \mathbf{I}$). The convergence of the method can be assured by the normalized potential energy fluctuation around a mean value after 400 HMC steps (Fig. \ref{fig:convergence}a), where we achieve an acceptance rate of $\sim 63\%$. After the burn-in phase (first 100 HMC steps), the phase space exploration becomes slower if the masses are not reduced, indicating the necessity for more HMC steps. This behaviour is less pronounced in the low data noise case due to the choice of maximum and minimum values to $\gamma_i(z)$.
Compared with a standard HMC experiment (dashed lines in Fig. \ref{fig:convergence}a), our strategy improves the convergence of the HMC method by decorrelating the samples (Fig. \ref{fig:convergence}b) as data noise increases.%

\begin{figure*}[htbp]
    \centering
    \includegraphics[width=6.8in]{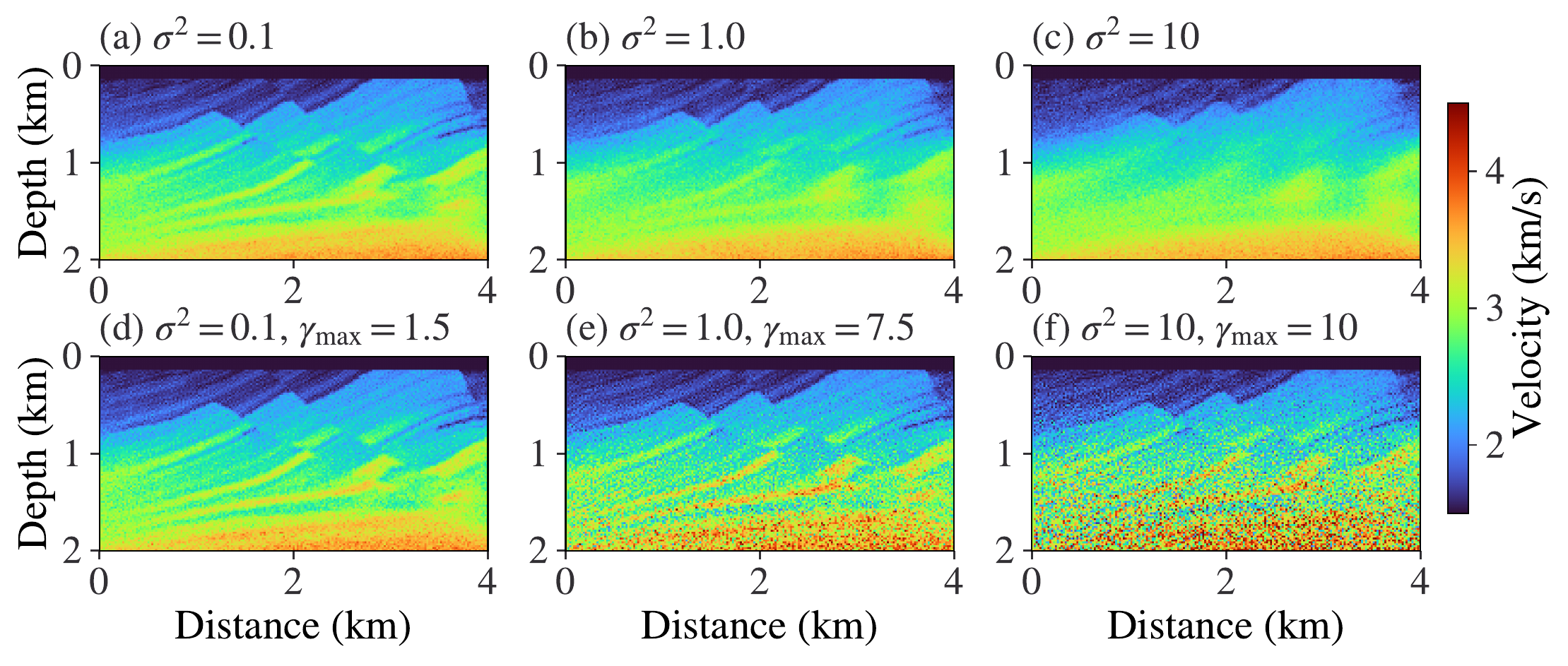}
    \caption{The maximum likelihood estimate (the most probable value for 300 samples) for each noise scenario, considering the standard (a), (b), (c) and the mass matrix proposed strategy (d), (e), (f). The tunning strategy proposed for the mass matrix can fast reconstruct reasonable seismic models compared to the conventional approach, therefore saving computational resources which is fundamental in FWI problems.}
    \label{fig:best_models}
\end{figure*}

The inversion results are illustrated in Fig. \ref{fig:best_models}, where we present the models that maximize the likelihood function (sample mode) and correspond to the solutions in deterministic inversion for each noise data case. We note that the HMC method following our tuning strategy (Fig. \ref{fig:best_models}d, e, f) is able to reconstruct the main features of the target model (Fig. \ref{fig:marm_target_init}a) faster than the conventional one (Fig. \ref{fig:best_models}a, b, c), mainly in the deep region ($z > 1$ km) which is poorly constrained by the data. This means that it is required more gradient calculations in the conventional approach making the problem more expensive. Therefore, we noted that a naive choice for the mass matrix turns this type of problem unfeasible to solve in a practical amount of time.

As expected, the resolution of the models are less affected when variance $\sigma^2$ of the residuals increases, at the price that the obtained models show a noisier aspect. In fact, the standard deviation $\sigma$ is interpreted as an effective searching radius of relevant models in data space $\mathbb{D}$. The size of this radius directly impacts the probability of sampling similar models, which motivates us to adapt the values used in the proposed strategy for tuning the matrix mass. Moreover, the relatively poor illumination at deeper regions of the model also can be related to the mean source frequency $f_0$ chosen to realize the experiments~\cite{curtis2020}. 

\begin{figure*}[htbp]
    \centering
    \includegraphics[width=6.8in]{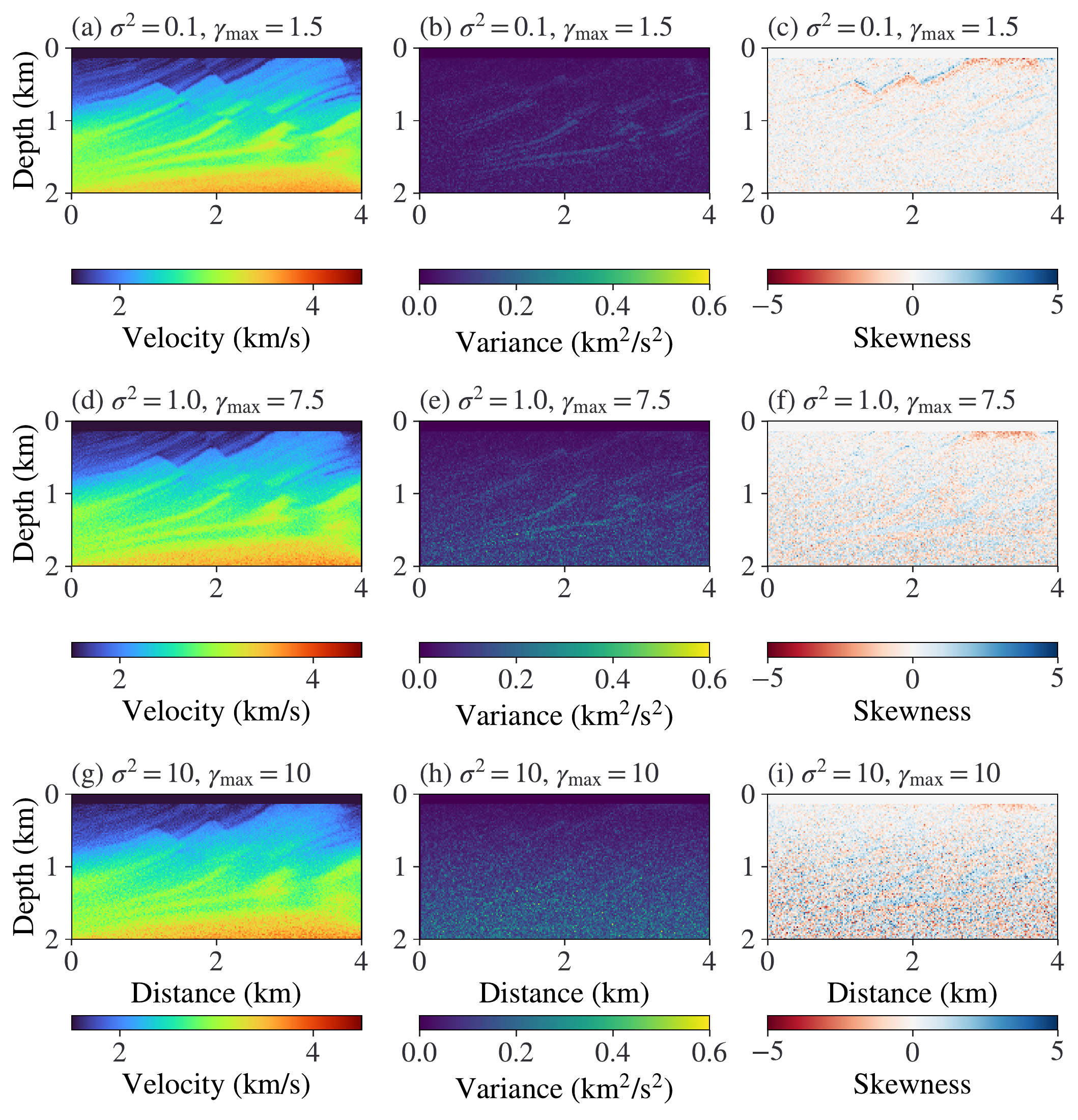}
    \caption{Summary of sample mean, variance and skewness to Marmousi target model (Fig. \ref{fig:marm_target_init}a) under different noise scenarios. As expected, the resolution of mean models (a, d, g) is influenced by the uncertainty in data. The skewness corresponding to the Marmousi model (c, f, i) shows that the mean are not sufficient to characterize the inversion and the model variance (b, e, h) indicates that the uncertainty is greater in discontinuities regions and increase in depth.}
    \label{fig:unc_analysis}
\end{figure*}

\begin{figure*}[htbp]
    \centering
    \includegraphics[width=6.8in]{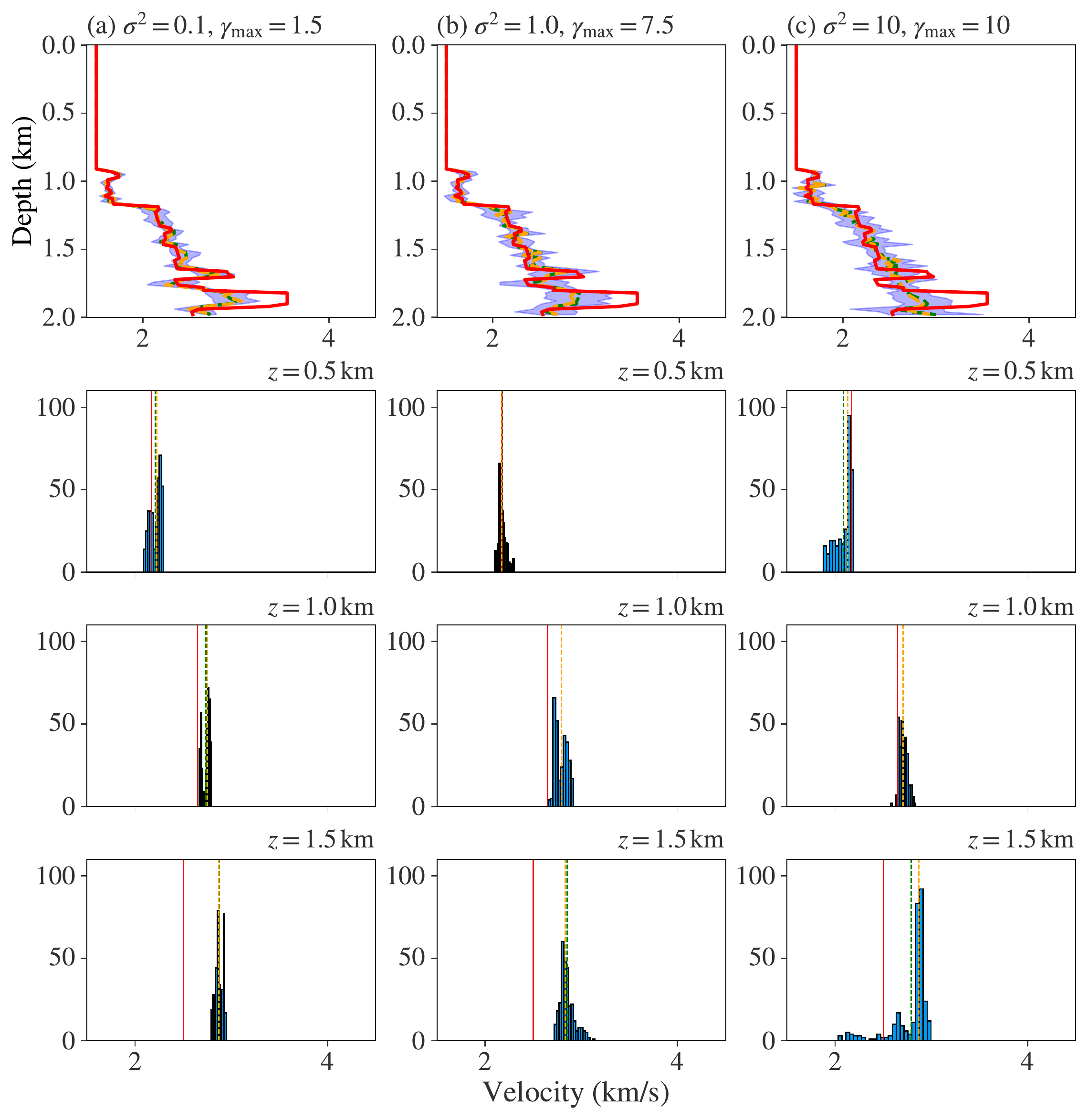}
    \caption{Uncertainty quantification in vertical profiles at horizontal distance of $x = 2$ km in (a) low, (b) medium and (c) high noise conditions. In each case, the histograms represents the model velocity distribution at depths: $z = 0.5 , 1.0,$ and $1.5$ km. The green, orange and red lines indicates the mean, median and true values, respectively. The blue shaded regions indicates a interval of two standard deviation centered in the mean values.}
    \label{fig:marginal_dist}
\end{figure*}

In addition, we assess the uncertainty in our FWI experiment by computing the mean, variance and skewness for the sample models under different variance scenarios (Fig. \ref{fig:unc_analysis}). We note that in  the shallow region  ($z < 1$ km) the mean velocity models (Fig. \ref{fig:unc_analysis}a, d, g) have a similar aspect to the target model (Fig. \ref{fig:marm_target_init}a), but only large-scale features are shown at deep regions ($z > 1$ km). The variance models (Fig. \ref{fig:unc_analysis}b, e, h) capture the Marmousi model discontinuities, probably due to the  sensitivity of the potential energy \eqref{misfit_function} to changes in traveltime along the model~\cite{Fichtner2018, curtis2021} and uncertainty loops~\cite{galletti2015}. However, in high variance scenario, this phenomenon is combined with the high model variance values of other regions, mainly of deeper regions. The histograms for particular model parameters (Fig. \ref{fig:marginal_dist}) show that uncertainty rises for increasing depth, which can be explained by the acquisition geometry nature of our seismic problem. We also verify an interchange (positive and negative values) in the skewness (Fig. \ref{fig:unc_analysis}c, f, i) along the anomalies of Marmousi model, which reveals the non-Gaussian behaviour of nonlinear inverse problems. Similarly to model variance, the non-Gaussianity increases and alternates it value with the depth (see for example Figure \ref{fig:marginal_dist}b and compare $z = 0.5$ with $z = 1.5$ km). This oscillation in the asymmetry of model distribution makes the most probable value (mode) greater or smaller than the mean value depending on the regions of the model and evidence that the mode is not sufficient to characterize the inversion. Similar results to the skewness values were reported in~\cite{izzatullah2021} using the Langevin dynamics. Therefore, sampling techniques based solely on gradient information and, even generalizations using Hessian information~\cite{sen2020}, do not provide a complete uncertainty quantification for our problem. 

\section{Conclusion}

In this work, we presented the application of Hamiltonian Monte Carlo (HMC) method to an acoustic Full-Waveform Inversion (FWI).  We also  proposed a new strategy of tuning the HMC matrix mass that improves the convergence maintaining high levels of acceptance rate. As expected to reflection experiments, the results show that the uncertainty increases with depth. We study in detail the dependence of variance $\sigma^2$ of the residuals in the HMC framework. The results shows a trade-off between image sharpness and accuracy. Small $\sigma^2$ produces sharp images but with poor accuracy in the velocity values, on the other side, large $\sigma^2$ have large uncertainty, the image is more blurred, but the mean velocity values are closer to the real ones. Additionally, the skewness values of our results demonstrates that statistical analysis based on Gaussian (or others symmetric) distributions has a limited meaning in reflection FWI. 

We developed an innovative strategy of tuning the HMC mass parameters with depth and algorithmic iteration devoted for reflection seismic problems. In order to capture large wavelength information in the beginning of the FWI process we start with a large HMC mass and decrease the mass with algorithm iteration to capture image details. Small masses in the beginning imply that the particles are less inert and can explore large phase space regions avoiding local minima and the cycle skipping effect. In contrast, large masses mean the particles visit smaller phase space regions  producing a sharp image. The proposed approach speed up the HMC convergence and open the doors to application in large scale problems. In a future work we plan to generalize our tuning strategy by changing masses for different iterations and use others prior information about the geological model.

\section*{Authors Contributions}

P. D. S. de Lima initiated the study, performed the simulations, analyzed the results, co-wrote, and reviewed the manuscript. M. S. Ferreira, G. Corso and J. M. de Araújo analyzed the results, co-wrote, and reviewed the manuscript. 

\begin{acknowledgments}

The authors gratefully acknowledge support from Shell Brasil through the “New Methods for Full Waveform Inversion” project at Universidade Federal do Rio Grande do Norte and the strategic importance of the support given by ANP through the R\&D levy regulation. We acknowledge NPAD/UFRN to allow us to use their computational resources. We thank CNPq (grant no. 313431/2018-3, 307907/2019-8) for funding.
\end{acknowledgments}

\bibliography{apssamp}

\end{document}